\documentclass[]{spie}  
\usepackage[]{graphicx}

\title{Characterization of LSST CCDs Using Realistic Images,
Before First Light}
\author{Andrew K. Bradshaw, Craig Lage, and J. Anthony Tyson
\skiplinehalf
University of California, Davis, 1 Shields Avenue, Davis, CA, USA\\
}

\authorinfo{Further author information: Send correspondence to A.K.B.: akbradshaw@ucdavis.edu}

\begin{document} 
\maketitle 

\begin{abstract}
The 3.2 gigapixel LSST camera, an array of 189 thick fully-depleted CCDs, will repeatedly image the southern sky and accomplish a wide variety of science goals. However, its trove of tens of billions of object images implies stringent requirements on systematic biases imprinted during shift-and-stare CCD observation. In order to correct for these biases which, without correction, violate requirements on weak lensing precision, we investigate CCD systematics using both simulations of charge transport as well as with a unique bench-top optical system matched to the LSST's fast f/1.2 beam. By illuminating single CCDs with realistic scenes of stars and galaxies and then analyzing these images with the LSST data management pipelines, we can characterize the survey's imaging performance well before the camera's first light. We present measurements of several CCD systematics under varying conditions in the laboratory, including the brightness-dependent broadening of star and galaxy images, charge transport anomalies in the silicon bulk as well as the edges, and serial deferred charge. Alongside these measurements, we also present the development and testing of physics-based models which inform corrections or mitigation strategies for these systematics. Optimization of the CCD survey operation under a variety of realistic observational conditions, including systematic effects from the optics, clocking, sky brightness, and image analysis, will be critical to achieve the LSST's goals of precision astronomy and cosmology.

\end{abstract}


\section{Introduction}
\label{sec:intro} 
The focal plane of the Large Synoptic Survey Telescope (LSST) will be tiled with an array of 189 charge-coupled devices (CCDs) totaling over 3 billion pixels. This gigantic camera, coupled with the LSST's uniquely wide and fast f/1.2 optics, will enable rapid surveying of the entire visible sky every few nights \cite{Ivezic2008arXiv}. This is due to the unprecedented etendue of the camera-telescope system. The statistical power of this data set will enable a wide variety of goals, from an inventory of our Solar System and Milky Way to investigating the nature of the accelerating Universe.  However, great statistical precision requires greater control of systematics. In order to achieve the diverse goals of the LSST, it is necessary to correct for an increasing number of systematic errors that are being discovered in the quest for sub-percent precision CCD observations \cite{Doherty2014SPIE}. 

Each of the LSST camera's 3.2 billion pixels are towers of silicon, $100 \mu m$ tall by $10 \mu m$ square, defined and read out by grids of potentials and ion implants at the surface and edges of each chip. In the silicon, electrons and holes liberated by the photoconversion process must find their way to the surface where they can be read out. However, on their way to the potential wells which comprise the pixel array, electrons can be diverted by many sources of stray electric fields. These transverse fields induce charge transport anomalies which  affect signal electrons in the bulk of the silicon, at its edges, or upon readout of the charge \cite{Stubbs2014JInst}. Each of them undermine the classical assumption of a CCD as a rectilinear grid of independent pixels that most analysis methods rely upon. Left uncorrected, these anomalies develop into systematic errors which bias the measured centroid, shape, and flux of object images, as well as any downstream measurements of the image properties of stars and galaxies which are needed to understand astrophysics and cosmology to sub-percent precision. In this report we will review a few of the most prominent systematic errors that we have characterized in the lab, including the brighter-fatter (BF) effect, astrometric distortions at the edge and in the bulk, and charge transport inefficiency. We present measurements of these effects on realistic stars and galaxies projected onto prototype LSST CCDs, which are then imaged, processed, and analyzed with state-of-the-art methods. Alongside these lab measurements we also present models of the microphysics of charge transport, inside of which we can ``turn'' some of the same ``knobs'' which are accessible to us in the lab, thereby informing the proposed methods of correction and making model predictions \cite{Lage2017JInst}.

\begin{figure}\includegraphics[width=.9\columnwidth]{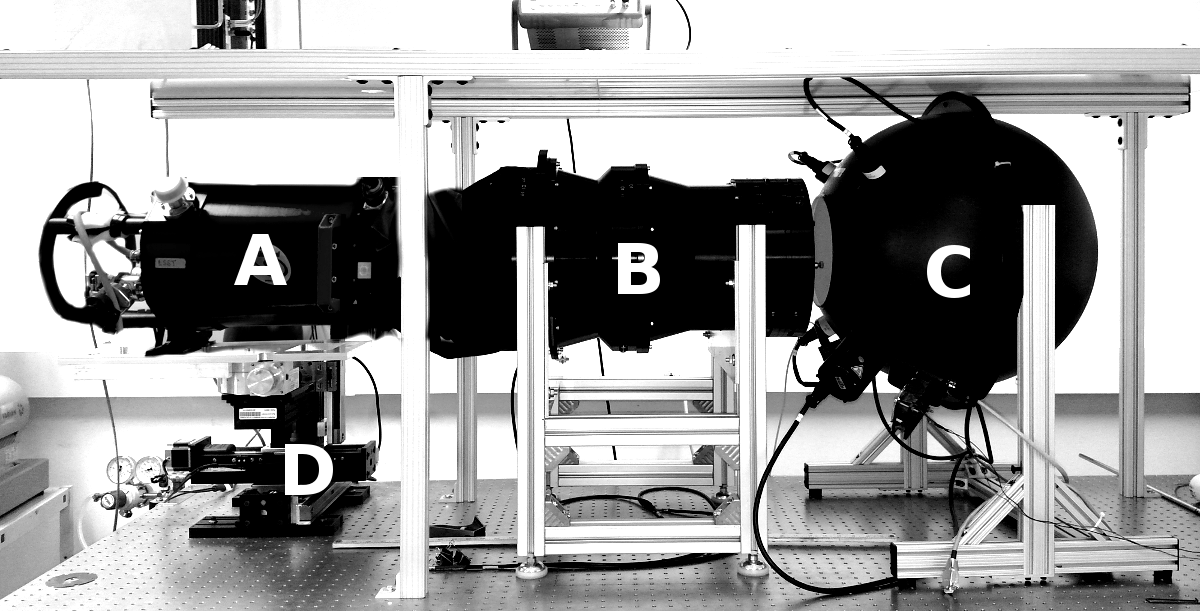}\centering
  \caption[Beam lab]{The LSST beam simulator at UC Davis. A: cryogenic Dewar with CCD mounted a few centimeters behind the input window. B: LSST f/1.2 beam simulator optics, where a spot mask with thousands of pinholes can be placed at the object plane. C: Illuminating sphere which scatters input light to provide uniform coverage of optics. D: Precision X/Y/Z stage upon which the CCD is mounted, enabling micron (1/10th of a pixel) dithering capability.}
  \label{fig:beamsim}
\end{figure}

In Section \ref{sec:setup} of this report we first describe the experimental setup which utilizes a fast f/1.2 beam simulator \cite{Tyson2014SPIE} to re-image realistic PSFs and galaxies onto prototype LSST CCDs (see Figure \ref{fig:beamsim}). We also briefly describe our image processing and analysis tools. In Section \ref{sec:bf} we describe the measurement and modeling of the BF effect, as well as our ongoing attempts at correction using a model in the literature. In Section \ref{sec:astrometry} we demonstrate the measurement and modeling of astrometric shifts due to readout electronics at the edges, and those due to tree rings in the bulk. In Section \ref{sec:cti} we present the measurement and impact of charge transport inefficiency (CTI), and deferred charge more generally, on the shapes of stars. We discuss these measurements and future plans in Section \ref{sec:discuss}, and conclude in Section \ref{sec:conclude}.

\section{Experimental setup and method} 
\label{sec:setup}
In the beam simulator laboratory shown in Figure \ref{fig:beamsim} we have full control of most observational parameters, including the brightness of the sky and its spectrum, the images of objects (ranging from stars and galaxies, to flats and targets), the full 6 degrees of freedom (sub-micron 3D translation + yaw/pitch/roll) to allow for precision focus and de-focus as necessary, temperature, clocking scheme, and all operational voltages. This comprehensive control allows an unprecedented level of testing under near-realistic conditions with a wide and fast beam capable of rapid characterization of imaging devices. The measurements made in the laboratory are complemented by a physics-based model of charge transport in silicon \cite{Lage2017JInst}, which provides excellent matches to laboratory tests under a variety of conditions. Using the data and simulations together enable the necessary physical understanding to develop algorithms to correct for an increasing number of systematic effects which can bias surveys of the sky.

To probe these systematics, we acquire and process data in a manner similar to that which will be used for the LSST survey. Namely, we take object images (e.g. of stars or galaxies) and calibrate them with standard techniques of gain estimation, bias subtraction, and flat fielding. Sequences of images can be acquired while varying the exposure time or brightness of light, dithering in the X/Y/Z direction, sky level, temperature, etc. These images therefore encompass the full spectrum of systematic effects associated with real observing, including optical PSFs with realistic aberrations and CCD systematics.

\begin{figure}\includegraphics[width=.9\columnwidth]{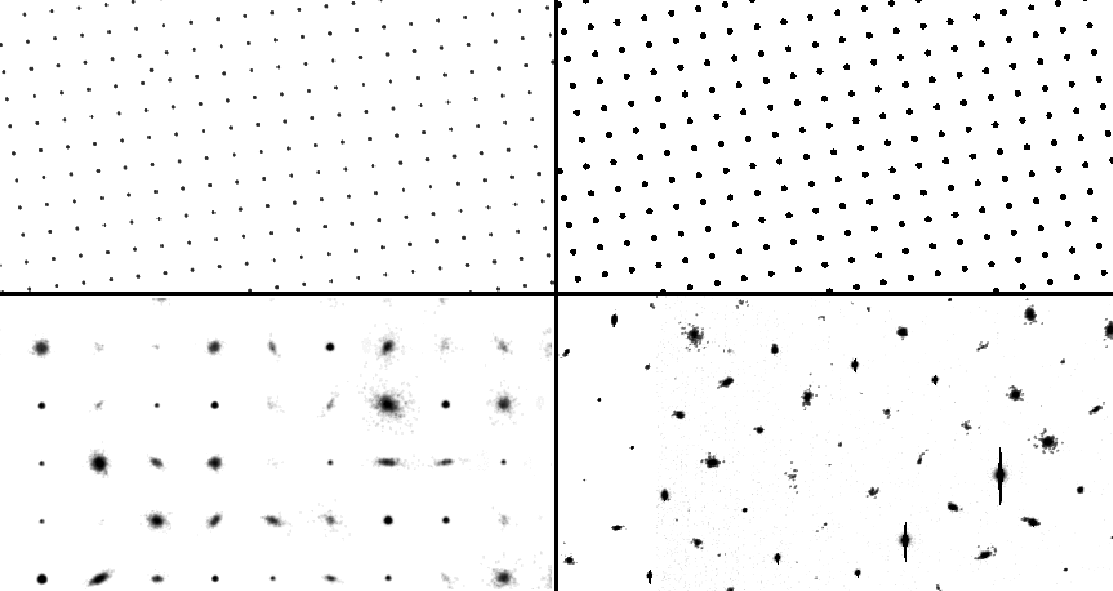}\centering
  \caption[The masks]{Screenshots (500 pixels wide) of the various photolithographic masks we currently have available for simulating LSST-like observing in the lab. Top left \& right: arrays of $3$ \& $30 \mu$m circular apertures to probe sub-pixel effects and median LSST seeing, respectively. Bottom left \& right: arrays of ``low'' and ``high''-resolution star/galaxy masks. These have stars of varying aperture sizes ($\sim$ PSF size \& magnitude) alongside galaxies with a sampling of realistic Sersic (index and effective radius) and ellipticity/orientation parameters. The effects of blooming on some stars with large apertures can be seen in the bottom right. Each mask can be rotated by a full 360 degrees.}
  \label{fig:themasks}
\end{figure}

In this report we focus on several types of object exposures: small $3\mu$m pinholes which are capable of probing sub-pixel effects, larger $30\mu$m pinholes which approximate the expected median LSST-like seeing distribution, and exposures containing both stars and galaxies with realistic parameters. A screenshot mosaic of these masks is shown in Figure \ref{fig:themasks}. In all cases, these objects are photolithographically printed on AR-coated quartz slides which are then placed into the object plane of the beam simulator: after the filter (chosen to be the R band in all of these tests) and before the optics. Exposure sequences used for this report were taken while varying either brightness (exposure time) or X/Y position (dithering). 

After acquisition, image processing proceeds as usual with steps for gain calibration, median bias and overscan subtraction, and flat fielding. Analysis of calibrated images can then be done in whatever fashion is desirable. Most pertinent are those methods which analyze the pixels in the same manner as the LSST, using the open-source software stack in development for this purpose \cite{Juric2015arXiv}. In particular, we use the measurements of centroid and shape parameters as probed by adaptive second moments \cite{Bernstein2002AJ} calculated on the pixelized image. These measurements, and the systematics which affect them, span a wide range of the LSST's scientific goals from asteroid detection to weak gravitational lensing. Additionally, we test multiple prototype CCDs with various known systematics, including those with high values of charge transport inefficiency (CTI) and tree rings. Testing of these flawed devices alongside science-grade ones can provide a useful baseline for modeling and correction.

\section{Brighter-fatter effect}
\label{sec:bf}
Chief among the systematic errors for modern precision surveys is the brighter-fatter (BF) effect, which biases the size and ellipticity of objects proportional to their brightness. More specifically, the transport of charge from generation to collection is distorted based upon the number of electrons already collected in the pixel wells. This brightness-dependent PSF effect has long been observed in the distributions of size versus magnitude of stars \cite{Astier2013A&A}, as well as a decrease in the spatial variance of flat fields taken at increasing intensities \cite{Downing2006SPIE}. Only recently has the BF effect become one of the most dominant sources of error in the science of weak gravitational lensing, where unbiased shape and size estimation of PSFs and galaxies is critical. Correction of this effect can take on several forms\cite{Antilogus2014JInst}, including redistribution of charge on a pixel by pixel basis \cite{Gruen2015JInst}, or through an effective re-mapping of the PSF and/or world coordinate system. We test the former method using an iterative kernel correction to object images \cite{Coulton2017arXiv}, finding varying degrees of success depending on the object parameters. We have also tested (but do not present here) the application of corrected PSFs for galaxy shape measurement using Sersic-fitting and a Gaussian mixture models, with further testing of shear measurement algorithms planned.

\subsection{BF effect on stars}
\begin{figure}\includegraphics[width=.9\columnwidth]{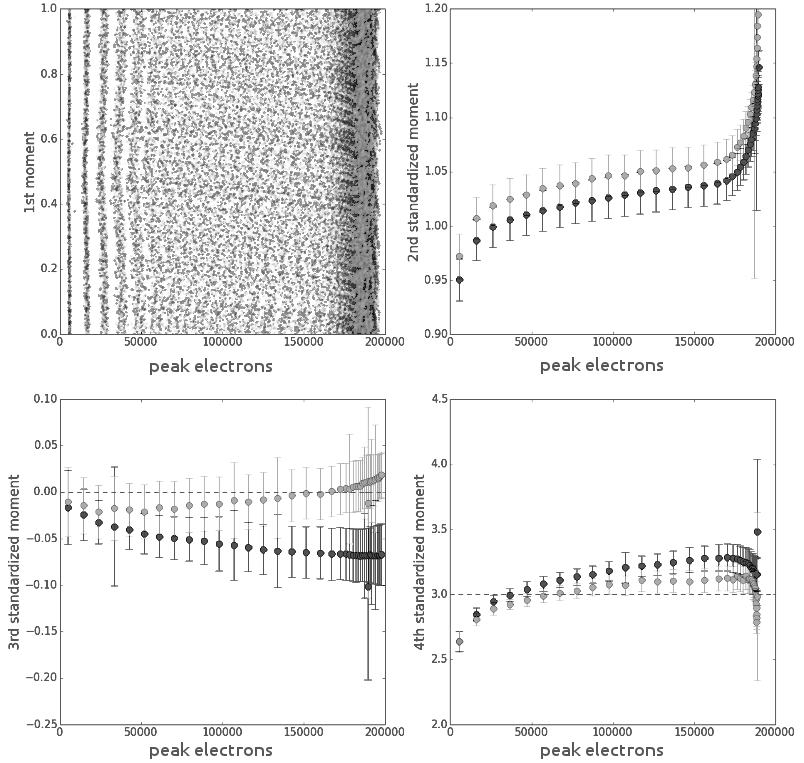}\centering
  \caption[Higher moments]{The brighter-fatter effect on 30 micron pinholes which approximate median LSST seeing. In each panel the method of moments is applied to tens of thousand of spots, providing image measurements in the X (black) and Y (grey) directions as brightness was increased. Error bars are computed from the dispersion of calculated moments in a given brightness bin, which have a large intrinsic spread due to pixelization. The first order (position or centroid) is shown in the top left, while second moment (width or  $\sigma$) in the top right. Skewness and kurtosis are shown in the bottom panels, indicating non-zero variations of moments beyond simple Gaussian spread induced by the brighter-fatter effect.}
  \label{fig:moments}
\end{figure}

The measurement of the distorted second moments measured on PSF images (30$\mu$m pinholes) as a function of brightness is shown in Figure \ref{fig:moments}. Thousands of spots were imaged and measured at increasing exposure levels to reveal the dependence of PSF shape on collected charges. The first moment, in the upper left, shows the sub-pixel centroid distribution of spots. There is a curious sign of preference for sub-pixel position, some of which is due to the dithering scheme but also some of which is dependent upon brightness. In the top right panel, the broadening of the second moment is the typical method for characterization of the BF effect, as it is most directly related to the shape parameters which are used in weak lensing. The linear rise of second moment around $10^4$ to $1.5\times 10^5$ peak electrons is the indicator of the effect, and has been largely captured with physical models of the charge transport \cite{Lage2017JInst}. At lower peak counts, the wings of the PSF do not have the necessary signal-to-noise to be detected and analyzed, thereby decreasing the measured width as shown; additionally, saturation of pixel wells can be seen as the sharp increase in second moment at high ($>1.5\times 10^5$) peak electron counts. 

The third and fourth moments as shown in the bottom left and right panels of Figure \ref{fig:moments}, which represent skewness and kurtosis, are also functions of brightness. In particular the kurtosis, which describes the profile’s tails and shoulders and is equal to 3 for a perfect Gaussian, shows a clear trend similar to the second moments. These measurements indicate that the brighter-fatter effect in CCDs is not simply a function of size. These higher-order effects of the brighter-fatter systematic error are important when considering the shapes of galaxies, which are typically non-Gaussian in their shape.

\subsection{BF effect on galaxies, and its correction}
As seen in the previous subsection, the brighter-fatter effect changes the width as well as the skewness (3rd) and kurtosis (4th) moments of nominally Gaussian spots. This non-zero response in the higher moments indicates the brighter-fatter effect has a more complicated consequences than a linear broadening, and that these consequences depend on the intrinsic profile of the object. Therefore, a model of how subsequent signal electrons will be deflected by the collected charge, and thus how a correction can be applied, will depend upon the natural un-aberrated, un-pixelized, and un-broadened shape of the object. Since this object profile is un-observable, iterative methods of correction must be applied. We apply one such method from the literature \cite{Coulton2017arXiv} which estimates incident count rates through successive redistribution of charge until convergence is reached. This model of charge redistribution, which is based on the expected deflection of electrons by the buildup of electric fields, takes the form of a kernel which can be applied to pixelized images of stars and galaxies in order to reconstruct their un-broadened profiles. This kernel can be derived from flat field images via an iterative method of accounting for lost variance, or via simulations of pixel boundary shifts \cite{Lage2017JInst}. We choose the latter kernel for these tests, as extraction from flats is complicated due to additional sources of variance which must be taken into account.

\begin{figure}\includegraphics[width=.8\columnwidth]{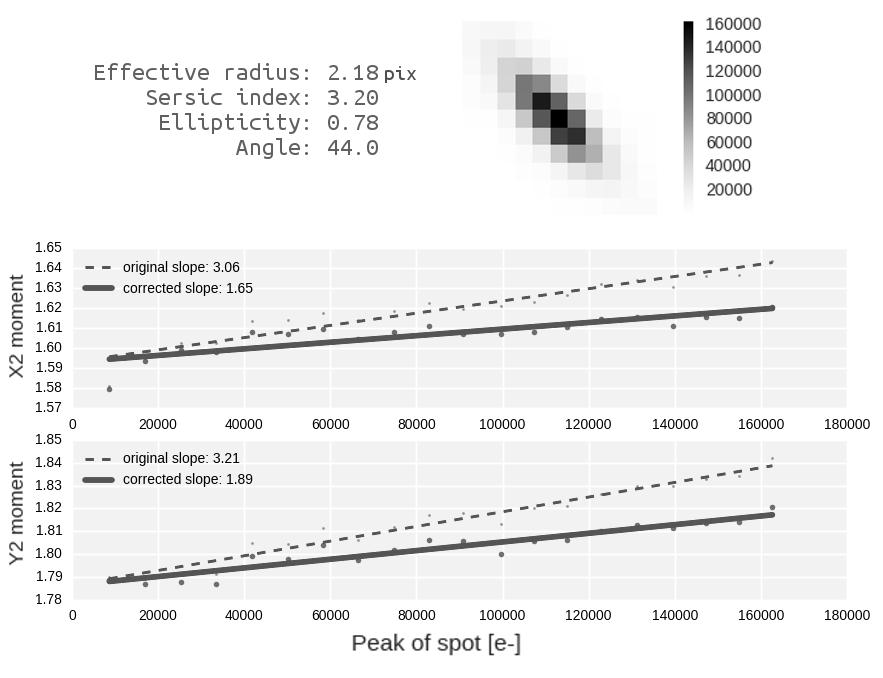}\centering
  \caption[Galaxy BF]{The brighter-fatter effect on one of the projected images of a galaxy with Sersic parameters and thumbnail shown in the top panel. In the middle and lower panel, adaptive second moments \cite{Bernstein2002AJ} in the X (middle) and Y (bottom) directions measured the galaxy's growth as a function of brightness, before and after application of the corrective kernel. Slopes are calculated via linear regression of second moments of spots which have peak counts $10^4 \geq e^-_{peak} \leq 1.5\times 10^5$. The kernel method can be seen to variously over- and under-correct the strength of the brighter-fatter effect which is dependent upon the intrinsic profile illuminating the pixels.}
  \label{fig:bf_gal}
\end{figure}

We test this kernel correction method on our array of stars and galaxies (as shown in the bottom half of Figure \ref{fig:themasks}) which contain numerous realistic galaxy profile shapes that were imaged with increasing exposure times. In Figure \ref{fig:bf_gal} we present just one of the hundreds of galaxies with which we performed brighter-fatter studies. The corrected and uncorrected brighter-fatter curves are shown in the figure as solid and dashed curves, and their measured slopes are shown in the legend. Each pair of X and Y measurements of image width as a function of brightness is performed using adaptive second moments \cite{Bernstein2002AJ} on the pixelized image. In the case shown, it can be seen that the kernel method indeed reduces the slope of the brighter-fatter effect, but doesn't correct the second moment entirely. We observe other combinations of intrinsic parameters such as Sersic indices, ellipticity, orientation, and brightness which appear to confound this first approximation to correction of the BF effect. We plan to probe these interesting irregularities through further development of the kernel and its application under more conditions, including new masks and rotations. 

\section{Edge \& Bulk Astrometric Shifts}
\label{sec:astrometry}
Beyond the brighter-fatter effect, there are at least two other charge transport systematics that redistribute incoming signal electrons by transverse electric fields: those at the edges due to the potentials on readout electronics, and those in the bulk induced by dopant variations i.e. tree rings. These astrometric distortions which are fixed (static) in their CCD location are particularly troublesome because of the pattern they imprint on every exposure. Correction of these effects previously involved division by a ``flat field'', however this method does not correct for the actual astrometric distortion induced by the transverse fields. Future corrections in development can take several forms beyond a simple cut on the image area that will satisfy the science requirements. A correction based upon the physics of the detector should be possible, and so we present tests of the astrometric error at the edge under varying observational conditions, matched to simulations, to aid in the development of a mitigation strategy.

\subsection{Edge distortion}
\begin{figure}\includegraphics[width=.9999\columnwidth]{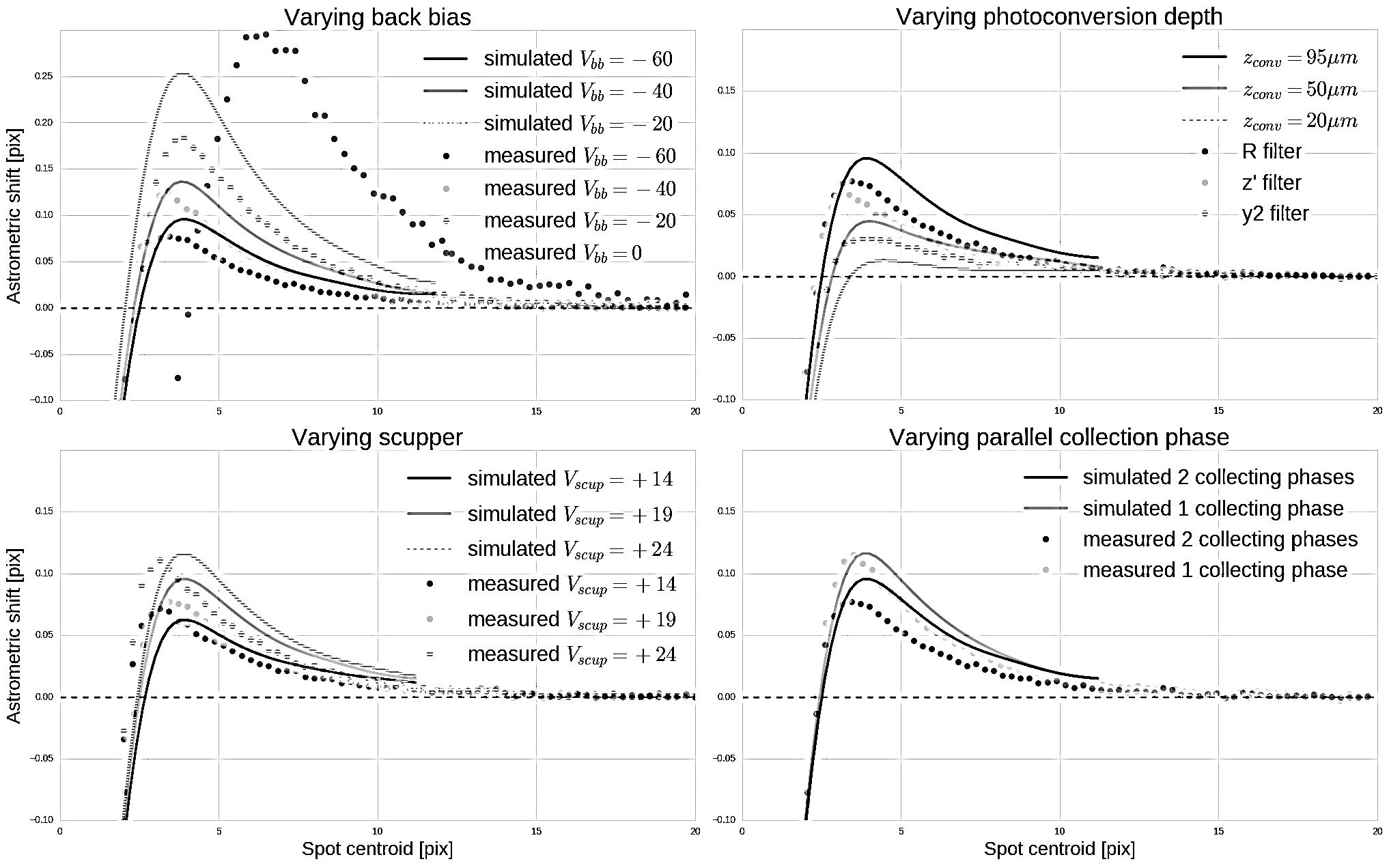}\centering
  \caption[Edge Effect]{Summary of the astrometric error at the edge as measured in the laboratory and simulated via Poisson’s equations. Operating voltages, clocking, and wavelength of light are modified in the laboratory and in the simulations and described in the text. Each variable has an effect on the edge distortion which is qualitatively similar to that seen in observations, indicating successful generalized modeling of charge transport.}
  \label{fig:edge_effect}
\end{figure}

In thick silicon CCDs, transverse electric fields sourced by the readout electronics measurably divert photoelectrons from purely vertical (i.e. parallel to the applied back bias field) transport into their pixels. For example, the so-called scupper ring is a feature implemented around the edges of the CCD package to protect the imaging array from any charges generated on the periphery. This metallization around the edge of the CCD is typically set at a positive potential that is high enough to induce measurable deviations of charge transport $\geq 10$ pixels into the bulk. Other metallizations and clocking lines on the periphery of each CCD can similarly divert signal electrons, and indeed the inclusion of each realistic detail in simulations has shown improvement in the matching to lab observations. Additionally, in the lab and in simulations we can vary multiple CCD operational voltages and clocking schemes as well as observe with differing PSF/object shape in other filters. Here we present observations, summarized in Figure \ref{fig:edge_effect} of tens of millions of $30\mu$m pinholes imaged under several different observational conditions in the lab. Four different experiments are shown in that figure, summarizing edge effect measurements which vary: 1) back bias voltage, 2) filter of observation (and thus photo-conversion depth), 3) scupper voltage, and 4) the charge collection scheme.

To measure the astrometric error in each of these schemes, thousands of pinholes are dithered approaching the CCD edge and the local displacement of spot centroids from their expected positions is calculated. Using this method of local deviation \cite{Bradshaw2015JInst} allows for a precision that is dependent only on the amount of data acquired (in this case, $\sigma_\mu (\Delta_{shift}) < .01 pix. = 100nm$). In each case, the outward diversion of photoelectrons is seen to increase in amplitude approaching the edge, peaking at $\sim 2$ pixels away. The astrometric error then quickly goes negative (retrograde), which is due to ordinary occultation of the optical spot at the edge and depends upon the intrinsic object size. In principle the centroid of these occulted (``masked'') spots can be modeled with more advanced methods, however, errors and assumptions made in the modeling process would only complicate the presentation of the measured centroid shift without adding physical insight.

In simulations, the photoelectron drift is captured by solving Poisson’s equations with realistic charges and potentials\cite{Lage2017JInst}. Collections of photoelectrons can be generated at known locations in the bulk and then allowed to drift according to the 3D electric fields induced by the readout electronics, pixel wells, and even bulk dopant variations. The resulting distorted pixel paths map pixel boundaries on ``the sky'' to those which are defined by the pixel array, and can be used to define distorted pixel boundaries which we use to pixelize a Gaussian profile similar to the one which we project onto the CCD in the lab. We then analyze the images in a manner identical to that used for lab data, measuring the difference between input and output centroids (including the occultation effect, by design). The simulations capture the general onset, shape, and amplitude of the astrometric error under varying observational conditions. This implies that the simulations provide generalized insight which can be tailored to any device and circumstance, and that the edge effect is a useful test case for physical modeling of imaging devices. This correspondence between the edge effect measurement and simulation independently validates the same methods which are used in the measurement of the brighter-wider effect previously presented in Section \ref{sec:bf}.

In our measurements and simulations, the back bias voltage has the largest impact on the amplitude of the edge effect. Operation of LSST CCDs currently uses a nominal potential of $V_{bb}=-60V$, which quickly sweeps photoelectrons to the pixel array. Lesser potentials (e.g. $-40, -20 V$) provide more time for photoelectrons to drift and thus more time for transverse potentials to affect charge transport, resulting in a larger astrometric shift seen in our measurements and simulations. Strangely, however, we find the laboratory that a zero back side bias still results in successful imaging (though with other unacceptable systematic errors), simulations of charge transport calculations with $V_{bb}=0$ do not converge. We believe this is due to the fact that the simulations assume a fully depleted silicon substrate, which is not the case with zero back-bias.

Varying the filter (wavelength) of observation also has an effect on the astrometric error, with larger astrometric error seen in bluer filters as shown in the upper right panel of Figure \ref{fig:edge_effect}. This wavelength dependence of the edge effect is due to longer transport times for photoelectrons generated higher in the bulk by photons with less penetration depth. Indeed, the reddest filter (y2) has the least astrometric error at the edge. Observations in the V or B bands, whose photoelectrons are almost entirely generated at surface of the CCD, have higher astrometric errors, though they are quantitatively similar to observations in the R band whose photons typically penetrate only a few microns into the bulk. 

Tests varying the scupper voltage indicate that a large portion of the astrometric error at the edge is induced by this protective potential, usually set to $\sim 20$ Volts. Making the scupper potential more positive increases the transverse force felt by the photoelectrons during their transport to the pixel well, such that varying this potential slightly by $\pm 5$ Volts has a measurable effect in the observations and simulations.

Finally, the parallel clocks which confine collected photoelectrons in their pixel rows can also have an effect on charge transport. Having two phases high during collection (the nominal operational condition for the LSST) slightly increases the average potential in the pixel array that is felt by photoelectrons during transport. Having only one phase high results in a less positive average potential in the bulk, implying longer charge transport times and more time to be affected by transverse edge potentials. This effect is measurable in the data and is also captured by the Poisson simulations.

\subsection{Tree Rings}
Another source of fixed-pattern astrometric distortion are the so-called tree rings seen in the silicon wafers used as CCD substrates. These tree rings are formed during growth of the crystal boule through dopant variations in the molten silicon. These $\sim 1\%$ impurity variations result in a fixed structure of electric fields that affect charge transport, which we have verified in simulations. These tree rings leave a visible pattern in flat fields and also induces a measurable astrometric error in PSFs of flat-fielded images. Since this effect is not related to QE, this raises the question of whether flat-fielding is a safe procedure for modern surveys \cite{Baumer2017PASP}. Each survey must answer this question individually, by taking into account the precision they aim to achieve and the characteristics of each detector. However, tree rings can bias cosmological parameters \cite{Okura2015JInst}, implying this astrometric error requires careful correction.

\begin{figure}\includegraphics[width=.95\columnwidth]{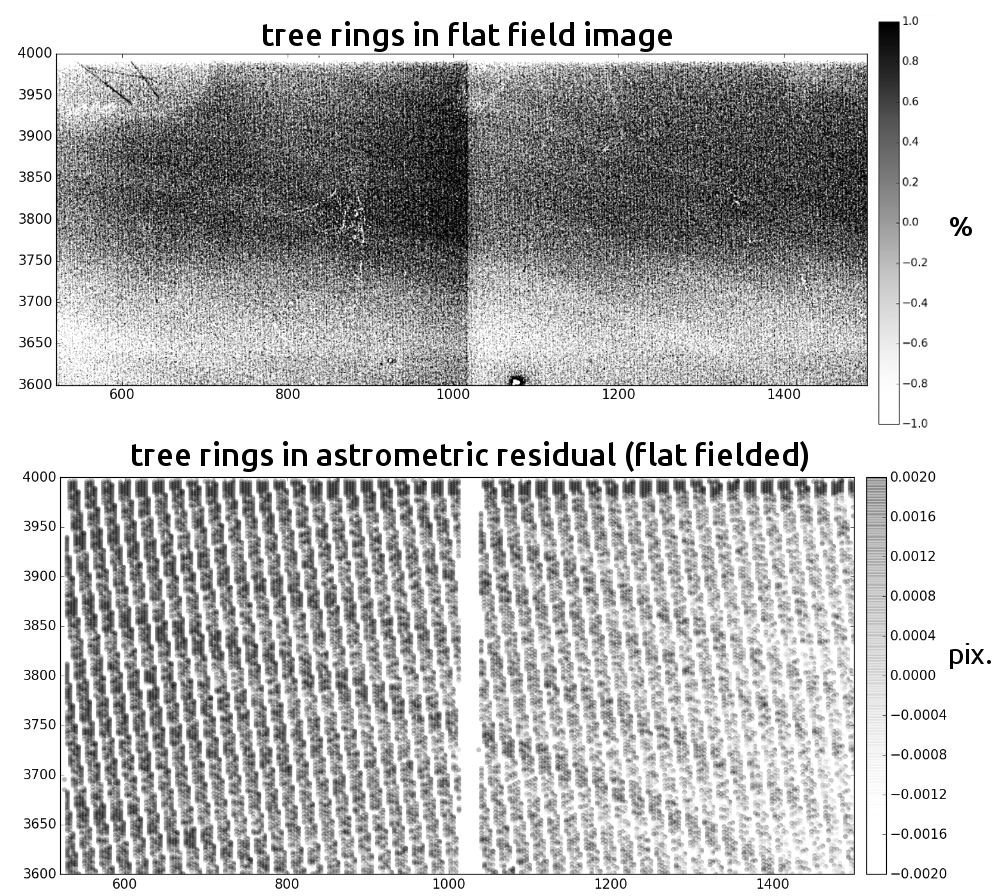}\centering
  \caption[Tree rings]{Sub-percent tree rings as observed in flat field flux (top) and as a thousandth of a pixel astrometric distortion (bottom). Flat field variations are expressed as percentage deviation from the mean by $\pm 1\%$. Astrometric shift in the Y centroid are presented in units of pixels from $\pm .002$. The astrometric error is calculated by differencing the expected and measured positions of thousands of spots which have been dithered across the frame, with gaps visible as white space in between segments (middle) as well as the zig-zag pattern resulting from the pinhole grid spacing. In addition to tree rings, edge effects in charge transport can also be seen as the sharp drop in the flat field flux and a sharp rise in the astrometric error at the top of each panel. Each image of the spots has been flat-fielded, indicating the astrometric error of both tree rings and edge effect are not purely a QE variation. }
  \label{fig:treerings}
\end{figure}

Using a prototype LSST CCD, we measure the effect of $\sim 0.5\%$-level tree rings in the lab using flat field images as well as the induced astrometric shift in $30\mu$m spots. These measurements, presented in Figure \ref{fig:treerings}, show a roughly 10-50X decrease of the tree ring amplitude and astrometric shift when compared to those observed in DECam \cite{Plazas2014JInst} and Pan-STARRS \cite{Magnier2018PASP} images. This reduction in tree ring amplitude is likely due to improvements in precise control of thermal and temporal variations during fabrication of the silicon boule, as well as strict acceptance criteria. The astrometric variation due to the $\sim 0.5\%$ RMS tree rings in this particular sensor induce an RMS astrometric shift of $\Delta=0.001$ pix. in PSFs similar to those expected in the median LSST seeing. 

\section{Deferred Charge}
\label{sec:cti}
Finally, we present measurements of the effect of apparent serial register charge transport inefficiency (CTI) on image shape parameters. Although CTI of ground-based detectors is usually much better than that in space (where charge traps can develop in the harsh environment), small inefficiencies in the rapid serial transport of charge can measurably bias the shapes of objects for a survey like the LSST. This investigation was motivated by a number of sensors discovered in LSST pre-testing that had unacceptably high levels of serial CTI as measured via the Extended Pixel Edge Response (EPER) technique in the overscan regions of flat field images. Testing these devices with varying levels of systematics, we are able to discern the effects of deferred charge. Like the tests presented before, these measurements inform physical models of the CCD which can be used in the development of correction algorithms. We briefly present one such model which indicates a correction may be possible.

In ideal flat field images, pixels which are serially read out after the last imaging column $Q_{last col}$ should have fluxes that immediately drop to the bias level of the amplifier. However, residual charge $Q_{deferred}$ is often seen in the first few overscan pixels at a measurable level. Traditionally, this deferred charge is attributed to traps in the silicon bulk which absorb charge during transfer and release it at a later time. Under this assumption, the deferred charge seen in the first overscan column of flat field images is the result of $N_{serial}$ inefficient transfers, and the ratio of deferred charge compared to the flat level is divided by this number of serial transfers.

However, there is growing evidence that the deferred charge seen in these EPER measurements of some LSST devices are not the result of traps in the device, but rather they are dominated by an imperfect response of the output transistor. To investigate this further, in the laboratory we make measurements of shape parameters using PSF-rich images and compare them to the measurements of CTI as quantified via the EPER technique on flat field on the same device. These shape measurements, though difficult to make due to the many layers of systematics, help to validate the proposed physical model of deferred charge.

\begin{figure}\includegraphics[width=.9\columnwidth]{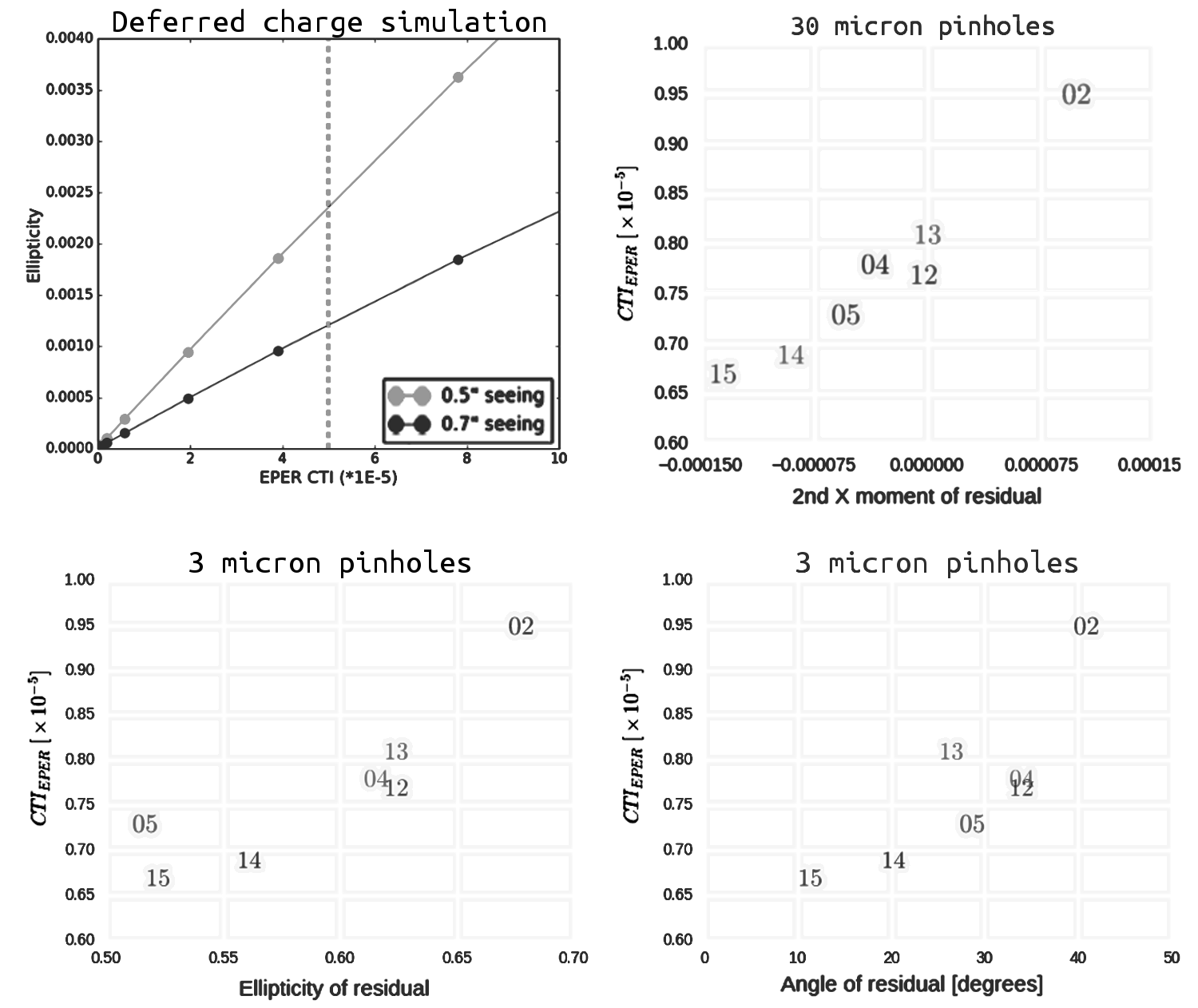}\centering
  \caption[CTI shapes]{Summary of the effect of deferred charge as quantified by $CTI_{EPER}$) on the shape of PSF-like light profiles. In the top left, we show the modeled effect of deferred charge on the ellipticity of PSFs, which increases as the deferred charge increases and as the PSF gets smaller. In all cases, the deferred charge is quantified via CTI as measured via the EPER technique on flats. In the top right and bottom panels, we present the effects as measured in the lab. The various measurements presented are given as residuals (as opposed to the absolute quantities available in simulations) due to experimental nuisances. In the top right, the effect of deferred charge on the 2nd moment in the X direction is detected using $30 \mu m$ spots which are similarly sampled to LSST median PSFs. In the bottom half, the measurements of ellipticity and angle of $3\mu m$ spots are shown for the same segments as labeled by their number. Both second moments and ellipticity can be seen to increase roughly linearly with $CTI_{EPER}$ on this CCD, in agreement with the model.}
  \label{fig:cti_shapes}
\end{figure}

The top left of Figure \ref{fig:cti_shapes} summarizes the proposed physical model with a range of CTI values. In this model, the effect of deferred charge on Gaussian spots was simulated by transferring an amount of charge proportional to CTI from each pixel to the next in the serial direction. For each value of deferred charge, the spot profiles were characterized using the same measurements as in the data: namely, windowed second image moments which enable ellipticity measurement of the spots. Note that for the X-axis (EPER CTI) we have taken the fraction of charge transferred and divided by 512, to be consistent with the way that the EPER CTI is measured. 

According to these simulations, the induced ellipticity on objects with typical levels of CTI ($\sim 10^{-6}$) is on the order of $10^{-5}$ or lower. The dotted vertical line in the top left panel shows a limit of $5.0 \times 10^{-5}$ which is is 10X the LSST's maximum CTI specification. This coherent PSF ellipticity would then get imprinted on 
each field of the sky every time it is re-observed, and the resulting shear bias would not average down like
root number of re-visits. If left uncorrected, their effect on the residual ellipticity correlation in the full 10-year survey would violate the requirement of $E=\langle \delta e \cdot \delta e \rangle \leq 2\times 10^{-7}$ (ellipticity of $0.00045$ per component) for objects separated by $\geq 5$ arcminutes. Therefore, testing of the CCDs destined for the LSST focal plane necessitates the quantification of deferred charge on ellipticity and verification of its model.

In the lab, we have investigated the effect of deferred charge on shape parameters using measurements of $3$ and $30 \mu$m pinhole arrays projected onto a CCD with segment-level $CTI_{EPER}$, measured to be $5\times 10^{-6}$ to $1\times 10^{-5}$ using the overscan of flat field images. The top right and bottom half of Figure \ref{fig:cti_shapes} show these measurements using several million $30 \mu$m (LSST-like PSFs) and $3 \mu$m spots. In all cases, there is a clear trend of increased second moment, ellipticity, and orientation angle with increased measures of deferred charge shown on the y-axis. In each case, the residual components of these measurements are shown, as the effect of deferred charge is not the sole systematic error in these real images. Optical aberrations as well as subpixel-level measurement systematics must be accounted for in order to detect the effect of deferred charge on a segment-by-segment basis. These experimental nuisances obscure the magnitude of the effect in absolute terms, and further interpretation and simulation is required beyond the detection of the effect presented here. In these figures, the segment number is plotted in lieu of symbols to aid in cross-plot visualization, and without error bars simply due to the complicated nature of error propagation in these sub-percent measurements. 
Low level PSF ellipticity systematics -- and the induced bias in weak gravitational lensing --  continue to challenge our analysis algorithms  and cosmological inferences.
\section{Discussion and Future Work}
\label{sec:discuss}
The measurements shown here represent a small fraction of the ongoing work involved in the quest for sub-percent precision CCD observing. Research into most of these effects is ongoing, and there are likely more systematics yet to be uncovered. These layers of systematics must be further investigated to better understand the physics behind them which aids in their removal. Therefore, more tests under a larger variety of conditions is warranted. 

For the brighter-fatter (BF) effect, optimization of operational voltages and performance of correction algorithms is a subject of great concern to the LSST, specifically the weak lensing community. We have shown that one available correction used in the literature \cite{Coulton2017arXiv} requires very careful implementation and understanding for application to all-sky datasets. For instance, a dependence of the BF slope (both corrected and uncorrected) was seen in the galaxy images we tested. Further testing of this is underway, with finer brightness sampling as well as mask orientation rotations. We also plan to deploy and test the performance of several shear measurement algorithms on our lab data using galaxy masks with known amounts of shear applied. Additionally, we will soon measure the BF effect on a different types of CCDs, applying the generality of the Poisson model to a new architecture of ion implants, voltages, and clocking schemes.  

Astrometric shifts at the edge and in the bulk are of less pressing concern due to their reduced amplitude, however they still warrant investigation. Here as in the BF effect, increased backside bias potential can mitigate these charge transport effects, but risk damage to the device. Laboratory optimization of operational voltages should be performed in the laboratory before on-sky measurements are made, in order to mitigate their effects and develop correction algorithms for real data.

We have also shown a detection of PSF aberration due to CTI (or more generally ``deferred charge'') on LSST-like PSFs as well as sub-pixel width Gaussians. Going beyond this detection, we developed a physical model which may aid in its mitigation through either sensor acceptance criteria, operational condition variation, or removal of remaining systematic effects in pixel processing.

In addition to the systematics presented here, we have investigated the effects of blooming as well as bloom-stops, the Binary Offset Effect \cite{Boone2018PASP}, pixelization biases, optical aberrations, and plan to investigate deblending and image coaddition techniques as well in order to aid in the development of correction algorithms. Further investigation of these systematics is still needed to quantify their effect on survey products after processing and correction methods have been applied.

\section{Conclusions}
\label{sec:conclude}
To maximize the scientific output of the survey, optimization of observing strategies and mitigation of systematic errors must be carried out before the survey begins. We present ongoing work in the lab to measure and correct for several prominent charge transport systematics using realistic images and modern processing and analysis tools. We quantify the brighter-fatter effect on realistic PSFs and galaxies, as well as test a correction method in development. We demonstrate the strong dependence of the charge transport systematics on operational conditions using the edge astrometric distortion, and successfully capture these variations in simulations. We also present the detection of astrometric distortion due to tree rings, and the effects of deferred charge ($\sim CTI$) are measured and compared to physics-based models. These laboratory measurements provide valuable insight that informs the development of correction algorithms. To properly correct for these systematic errors, new techniques must be developed to sequentially peel back the layers of physical effects which obscure the truth in modern surveys of the sky.

\acknowledgments      
We gratefully acknowledge Kirk Gilmore for CCD Dewar setup, Sven Herrmann and Claire Juramy for valuable discussions, and Dan Polin for lab assistance. Software used: LSST Data Management Science Pipelines Software \cite{Juric2015arXiv}. This material is based upon work supported in part by the NSF through Cooperative Agreement 1258333 managed by the Association of Universities for Research in Astronomy (AURA), and the DOE under Contract No. DE-AC02-76SF00515 with the SLAC National Accelerator Laboratory. Additional LSST funding comes from private donations, grants to universities, and in-kind support from LSSTC Institutional Members. Financial  support  from  DOE  grant DE-SC0009999 and Heising-Simons Foundation grant 2015-106 are gratefully acknowledged.


\bibliography{report}   

\begin{thebibliography}{10}

\bibitem{Ivezic2008arXiv}
{Ivezi{\'c}}, {\v Z}., {Kahn}, S.~M., {Tyson}, J.~A., {Abel}, B., {Acosta}, E.,
  {Allsman}, R., {Alonso}, D., {AlSayyad}, Y., {Anderson}, S.~F., {Andrew}, J.,
  and et~al., ``{LSST: from Science Drivers to Reference Design and Anticipated
  Data Products},'' {\em ArXiv e-prints}~{\bf 0805.2366} (2008).

\bibitem{Doherty2014SPIE}
{Doherty}, P.~E., {Antilogus}, P., {Astier}, P., {Chiang}, J., {Gilmore},
  D.~K., {Guyonnet}, A., {Huang}, D., {Kelly}, H., {Kotov}, I., {Kubanek}, P.,
  {Nomerotski}, A., {O'Connor}, P., {Rasmussen}, A., {Riot}, V.~J., {Stubbs},
  C.~W., {Takacs}, P., {Tyson}, J.~A., and {Vetter}, K., ``{Electro-optical
  testing of fully depleted CCD image sensors for the Large Synoptic Survey
  Telescope camera},'' in [{\em High Energy, Optical, and Infrared Detectors
  for Astronomy VI}{\nolinebreak\hspace{0.1em}]},  {\em Proc. SPIE} {\bf 9154},
   915418 (2014).

\bibitem{Stubbs2014JInst}
{Stubbs}, C.~W., ``{Precision astronomy with imperfect fully depleted CCDs - an
  introduction and a suggested lexicon},'' {\em Journal of
  Instrumentation}~{\bf 9},  C03032 (2014).

\bibitem{Lage2017JInst}
{Lage}, C., {Bradshaw}, A., and {Tyson}, J.~A., ``{Measurements and simulations
  of the brighter-fatter effect in CCD sensors},'' {\em Journal of
  Instrumentation}~{\bf 12},  C03091 (2017).

\bibitem{Tyson2014SPIE}
{Tyson}, J.~A., {Sasian}, J., {Gilmore}, K., {Bradshaw}, A., {Claver}, C.,
  {Klint}, M., {Muller}, G., {Poczulp}, G., and {Resseguie}, E., ``{LSST
  optical beam simulator},'' in [{\em High Energy, Optical, and Infrared
  Detectors for Astronomy VI}{\nolinebreak\hspace{0.1em}]},  {\em Proc. SPIE}
  {\bf 9154},  915415 (2014).

\bibitem{Juric2015arXiv}
{Juri{\'c}}, M., {Kantor}, J., {Lim}, K., {Lupton}, R.~H., {Dubois-Felsmann},
  G., {Jenness}, T., {Axelrod}, T.~S., and {LSST Project}, ``{The LSST Data
  Management System},'' {\em ArXiv e-prints}~{\bf 1512.07914} (2015).

\bibitem{Bernstein2002AJ}
{Bernstein}, G.~M. and {Jarvis}, M., ``{Shapes and Shears, Stars and Smears:
  Optimal Measurements for Weak Lensing},'' {\em AJ}~{\bf 123},  583--618 (Feb.
  2002).

\bibitem{Astier2013A&A}
{Astier}, P., {El Hage}, P., {Guy}, J., {Hardin}, D., {Betoule}, M., {Fabbro},
  S., {Fourmanoit}, N., {Pain}, R., and {Regnault}, N., ``{Photometry of
  supernovae in an image series: methods and application to the SuperNova
  Legacy Survey (SNLS)},'' {\em A\& A}~{\bf 557},  A55 (2013).

\bibitem{Downing2006SPIE}
{Downing}, M., {Baade}, D., {Sinclaire}, P., {Deiries}, S., and {Christen}, F.,
  ``{CCD riddle: a) signal vs time: linear; b) signal vs variance:
  non-linear},'' in [{\em High Energy, Optical, and Infrared Detectors for
  Astronomy II}{\nolinebreak\hspace{0.1em}]},  {\em Proc. SPIE} {\bf 6276},
  627609 (2006).

\bibitem{Antilogus2014JInst}
{Antilogus}, P., {Astier}, P., {Doherty}, P., {Guyonnet}, A., and {Regnault},
  N., ``{The brighter-fatter effect and pixel correlations in CCD sensors},''
  {\em Journal of Instrumentation}~{\bf 9},  C03048 (2014).

\bibitem{Gruen2015JInst}
{Gruen}, D., {Bernstein}, G.~M., {Jarvis}, M., {Rowe}, B., {Vikram}, V.,
  {Plazas}, A.~A., and {Seitz}, S., ``{Characterization and correction of
  charge-induced pixel shifts in DECam},'' {\em Journal of
  Instrumentation}~{\bf 10},  C05032 (2015).

\bibitem{Coulton2017arXiv}
{Coulton}, W.~R., {Armstrong}, R., {Smith}, K.~M., {Lupton}, R.~H., and
  {Spergel}, D.~N., ``{Exploring the brighter fatter effect with the Hyper
  Suprime-Cam},'' {\em ArXiv e-prints}~{\bf 1711.06273} (2017).

\bibitem{Bradshaw2015JInst}
{Bradshaw}, A., {Lage}, C., {Resseguie}, E., and {Tyson}, J.~A., ``{Mapping
  charge transport effects in thick CCDs with a dithered array of 40,000
  stars},'' {\em Journal of Instrumentation}~{\bf 10},  C04034 (2015).

\bibitem{Baumer2017PASP}
{Baumer}, M., {Davis}, C.~P., and {Roodman}, A., ``{Is Flat fielding Safe for
  Precision CCD Astronomy?},'' {\em PASP}~{\bf 129}(8),  084502 (2017).

\bibitem{Okura2015JInst}
{Okura}, Y., {Plazas}, A.~A., {May}, M., and {Tamagawa}, T., ``{Spurious shear
  induced by the tree rings of the LSST CCDs},'' {\em Journal of
  Instrumentation}~{\bf 10},  C08010 (2015).

\bibitem{Plazas2014JInst}
{Plazas}, A.~A., {Bernstein}, G.~M., and {Sheldon}, E.~S., ``{Transverse
  electric fields' effects in the Dark Energy Camera CCDs},'' {\em Journal of
  Instrumentation}~{\bf 9},  C04001 (2014).

\bibitem{Magnier2018PASP}
{Magnier}, E.~A., {Tonry}, J.~L., {Finkbeiner}, D., {Schlafly}, E., {Burgett},
  W.~S., {Chambers}, K.~C., {Flewelling}, H.~A., {Hodapp}, K.~W., {Kaiser}, N.,
  {Kudritzki}, R.-P., {Metcalfe}, N., {Wainscoat}, R.~J., and {Waters}, C.~Z.,
  ``{Charge Diffusion Variations in Pan-STARRS1 CCDs},'' {\em PASP}~{\bf
  130}(6),  065002 (2018).

\bibitem{Boone2018PASP}
{Boone}, K., {Aldering}, G., {Copin}, Y., {Dixon}, S., {Domagalski}, R.~S.,
  {Gangler}, E., {Pecontal}, E., and {Perlmutter}, S., ``{A Binary Offset
  Effect in CCD Readout and Its Impact on Astronomical Data},'' {\em PASP}~{\bf
  130}(6),  064504 (2018).

\end{thebibliography}
\bibliographystyle{spiebib}   

\end{document}